\documentclass{aa}
\usepackage{graphicx}
\usepackage{txfonts}
\usepackage{amsmath}
\usepackage{amsfonts}
\usepackage{amssymb}
\usepackage{color,hyperref}
\hypersetup{colorlinks,breaklinks,
            linkcolor=blue,urlcolor=blue,
            anchorcolor=blue,citecolor=blue}

\newcommand{\mrm}[1]{\mathrm{#1}}
\newcommand{\mbb}[1]{\mathbf{#1}}

\newcommand{\dx}{\ensuremath{\mathrm{\,d}}}

\begin{document}

\title{\texttt{Angpow}: a software for the fast computation of accurate tomographic power spectra}
\author{J.-E. Campagne \and J. Neveu \and S. Plaszczynski}
\institute{LAL, Univ. Paris-Sud, CNRS/IN2P3, Universit\'e Paris-Saclay, Orsay, France \\ \email{campagne@lal.in2p3.fr}}

 \date{Received 5 January 2017 / Accepted 27 April 2017}

\abstract{}{The statistical distribution of galaxies is a powerful probe to constrain cosmological models and gravity. In particular, the matter power spectrum $P(k)$ provides information about the cosmological distance evolution and the galaxy clustering. However the building of $P(k)$ from galaxy catalogs requires a cosmological model to convert angles on the sky and redshifts into distances, which leads to difficulties when comparing data with predicted $P(k)$ from other cosmological models, and for photometric surveys like the Large Synoptic Survey Telescope (LSST). The angular power spectrum $C_\ell(z_1,z_2)$ between two bins located at redshift $z_1$ and $z_2$ contains the same information as the matter power spectrum, and is free from any cosmological assumption, but the prediction of $C_\ell(z_1,z_2)$ from $P(k)$ is a costly computation when performed precisely.}{The \texttt{Angpow} software aims at quickly and accurately computing the auto ($z_1=z_2$) and cross ($z_1\neq z_2$) angular power spectra between redshift bins. We describe the developed algorithm based on developments on the Chebyshev polynomial basis and on the Clenshaw-Curtis quadrature method. We validate the results with other codes, and benchmark the performance.  }{\texttt{Angpow} is flexible and can handle any user-defined power spectra, transfer functions, and redshift selection windows. The code is fast enough to be embedded inside programs exploring large cosmological parameter spaces through the $C_\ell(z_1,z_2)$ comparison with data. We emphasize that the Limber's approximation, often used to speed up the computation, gives incorrect $C_\ell$ values for cross-correlations. The {\tt C++} code is available from \href{https://gitlab.in2p3.fr/campagne/AngPow}{https://gitlab.in2p3.fr/campagne/AngPow}.}{}

\maketitle

\section{Introduction}
Cosmology is entering the era of wide and deep surveys of galaxies, such as, for example, with the Dark Energy Spectroscopic Instrument (DESI) \citep{2013arXiv1308.0847L},  the Large Synoptic Survey Telescope (LSST) \citep{2008arXiv0805.2366I}, and the \textit{Euclid} satellite  \citep{2011arXiv1110.3193L}, in
order to investigate the mechanisms of cosmic acceleration \citep[for
a review see][]{Weinberg2013}.
Cosmological models can be tested, that is, compared against actual measurements, by studying the statistical
properties of galaxy clustering.
Several methods exist for this, the most classical ones computing correlations in real \citep{Landy1993}
or Fourier space \citep{Feldman1994}.
However, for wider and deeper surveys, one may also try to condense the
clustering information into redshift bins ("shells") and compute the auto- and cross-correlations between redshift shells \citep{Asorey2012}. This is known as tomography, and allows for a more precise
understanding of potential systematic errors in different redshift regions. Several studies compare
the merits of this kind of approach with the more classical ones
\citep{Asorey2012,Asorey2014,DiDio2014,Nicola2014,Lanusse2015} and try to optimize the binning to keep most of
the cosmological information.
All previous studies have been based on the Fisher formalism, which considers
observables as Gaussian; unrepresentative, however, of real life data.

To go on further and prepare the future tomographic analyses, one needs to implement a full pipeline
and test the method with, for instance, a Monte-Carlo Markov Chain (MCMC) exploration of the cosmological parameter space. 
But there is a technical bottleneck; running a typical MCMC algorithm in cosmology is already very lengthy and requires computing typically a few $10^5$ models. Each model is the result of a
numerical code that solves the cosmological equations (known as
"Boltzmann solvers"), such as {\tt
  CLASS}\footnote{\href{http://class-code.net}{http://class-code.net}}
\citep{classII}, which takes
typically 5-10 s on eight cores. Today  this amounts to several days of computation.

For a tomographic method, one further needs to transform the output of
the Boltzmann solver, the matter power spectrum, 
into the observable space, represented as $C_\ell(z_i,z_j)$ angular power spectra between  two redshift shells located at $z_i$ and $z_j$  . This transformation is numerically challenging because of overlapping integrals between very oscillating spherical Bessel functions.

One then often makes use of the Limber's approximation, which essentially replaces the Bessel functions by a single Dirac value. However, as we show here, this leads to a poor approximation for auto-correlations and may
even be incorrect for cross-correlations, since it cannot capture any anti-correlation.

We therefore address here the issue of accurately and quickly computing the integrals
required to derive the correlations between tomographic bins.
Our goal, in computational terms, is that this computation be faster than one typical
Boltzmann code computation time, that is, essentially at or below the 1s
level (on eight cores).
Another aspect of this work is to provide a stand-alone library that offers a generic interface where the user can plug any matter power spectrum. This is a different approach from {\tt CLASSgal} \citep{ClassGal} which also provides some theoretical computations related to tomography that are deeply rooted within the {\tt CLASS} software.

The integrals defining the $C_\ell(z_i,z_j)$ angular power spectra are introduced in section \ref{seq-pos}. In section \ref{seq-math} we detail the algorithm implemented in \verb|Angpow,| while we address some numerical tests in section \ref{sec-num-test}. Section~\ref{sec:code-design} provides insight into the code design and we conclude in section~\ref{sec-summary}. 
\section{The position of the problem}\label{seq-pos}
  
Our aim is to compute the angular over density power spectrum  $C_{\ell}(z_1, z_2)$ as a cross-correlation between two $z$-shells with mean values ($z_1, z_2$) and also the auto-correlation $C_{\ell}(z_1)$ with $z_1 = z_2$,  taking into account, in both cases, possible redshift selection functions. Following notations of reference \citep{ClassGal},  for a couple of redshift
$(z_1, z_2)$ one computes $C_{\ell}(z_1, z_2)$ according to
\begin{equation}
C_{\ell}(z_1, z_2) = \frac{2}{\pi} \int_0^\infty \dx k\ k^2\ P(k) \Delta_{\ell}(z_1, k)\Delta_{\ell}(z_2, k)
\label{eq-cl-def}
,\end{equation}
with on one hand $P(k)$ the non-normalized primordial power spectrum, and on the other hand, $\Delta_{\ell}(z, k),$ a general function used to describe physical processes down to redshift $z$ \citep{ClassGal, 2011PhRvD..84f3505B}. At the lowest order, $\Delta_{\ell}(z, k)$ can be expressed as the product of the bias $b$ and a growth factor $D(z,k)$ to account for matter density contribution:
\begin{equation}
\Delta^{\mrm{mat.}}_{\ell}(z, k) = b D(z,k) j_{\ell}(k\,r(z))
\label{eq-DeltaFunc-def}
,\end{equation} 
where $j_\ell(x)$ is a first kind spherical Bessel function of parameter $\ell$, and $r(z)$ is the radial comoving distance of the shell located at redshift $z$. 

For thick redshift shells, one introduces two normalized redshift selection functions $W_1(z;z_1,\sigma_1)$ and $W_2(z^\prime;z_2,\sigma_2)$  around $z_1$ and $z_2$ with typical width $\sigma_1$ and $\sigma_2$, respectively. Then, one extends equation \ref{eq-cl-def} by
\begin{equation}
\begin{split}
C^{\mrm{thick}}_{\ell}&(z_1, z_2;\sigma_1, \sigma_2) \\&=  \frac{2}{\pi} \iint_0^\infty \dx z \dx z^\prime \ W_1(z; z_1, \sigma_1) W_2(z^\prime; z_2, \sigma_2)\\
&\phantom{\iint_0^\infty \mathrm{d} z \mathrm{d} z^\prime} \times \int_0^\infty \dx k\ k^2\ P(k)  \Delta_{\ell}(z, k)\Delta_{\ell}(z^\prime, k) 
\end{split}
.\end{equation}
It is convenient to introduce the auxiliary  function justified in the following section
\begin{equation}
\begin{split}
f_\ell(z,k) &\equiv \sqrt{\frac{2}{\pi}}\  k \sqrt{P(k)} \ \Delta_{\ell}(z, k) \\
&=  \sqrt{\frac{2}{\pi}}\  k \sqrt{P(k)} \left\{ \vphantom{\sqrt{P(k)}}b D(z,k) j_{\ell}(k\,r(z)) + \dots\right\} \\
&= \sqrt{\frac{2}{\pi}}\  k \sqrt{P(k)D(z,k)^2} \left\{ \vphantom{\sqrt{P(k)}} bj_{\ell}(k\,r(z)) + \dots\right\} \\
&\equiv  \sqrt{\frac{2}{\pi}}\  k \sqrt{P(k,z)}\ \widetilde{\Delta}_\ell(z,k)
\end{split}
\label{eq-fell-func}
,\end{equation}
where we have used the factorization of the growth factor $D(k,z)$ from the matter density contribution to introduce the notation $P(k,z)$ and $\widetilde{\Delta}_\ell(z,k)$. The dots signify that  other contributions may be introduced as the redshift distortions and lensing effects that we ignore here for simplicity. Then, 
\begin{equation}
\begin{split}
C^{\mathrm{thick}}_{\ell}&(z_1, z_2;\sigma_1, \sigma_2)\\& = \iint_0^\infty \mathrm{d} z \mathrm{d} z^\prime \ W_1(z; z_1, \sigma_1) W_2(z^\prime; z_2, \sigma_2)\\
&\phantom{\iint_0^\infty \mathrm{d} z \mathrm{d} z^\prime} \times \int_0^\infty \mathrm{d} k\ f_{\ell}(z, k) f_{\ell}(z^\prime, k). 
\end{split}
\label{eq-clz1z2-obs}
\end{equation}

The auto-correlation is a
particular case where the redshift selection function $W_2(z^\prime;z_2,\sigma_2)$ is reduced to $W_1(z^\prime;z_1,\sigma_1)$ and we can use a single $W$ function, which leads to
\begin{equation}
\begin{split}
C^{\mrm{thick}}_{\ell}(z_1&;\sigma_1) = \iint_0^\infty \dx z\ \dx z^\prime\  W(z; z_1, \sigma_1) W(z^\prime; z_1, \sigma_1)\\
&\phantom{\iint_0^\infty \mathrm{d} z \mathrm{d} z^\prime} \times \int_0^\infty \mathrm{d} k\ f_{\ell}(z, k) f_{\ell}(z^\prime, k). 
  \label{eq-clz1-obs}
 \end{split}
\end{equation} 
To account for infinite redshift precision at $z=z_1$, the use of a Dirac selection function for $W$ yields
\begin{equation}
C^{\delta}_{\ell}(z_1) =  \frac{2}{\pi} 
 \int_0^\infty \dx k\ k^2 P(k,z_1) \widetilde{\Delta}^2_{\ell}(z_1, k).
  \label{eq-clz1-delta}
\end{equation}
\verb|Angpow| is designed to efficiently compute these $C_\ell$ expressions once one provides access to the power spectra $P(k,z)$, the $ \tilde{\Delta}_{\ell}(z, k)$ extra function, and the cosmological distance $r(z)$. To simplify the notation, we do not write the width $\sigma$ of the selection functions if not explicitly needed. 

\section{A brief description of the computational algorithm}
\label{seq-math}
The redshift integral computations of Eq.~\ref{eq-clz1z2-obs}  can be conducted in practice inside the rectangle $ [z_{1\mrm{min}},z_{1\mrm{max}}] \times [z_{2\mrm{min}},z_{2\mrm{max}}]$ given by the $W$ selection functions using  a Cartesian product of a one-dimensional (1D) quadrature 
 defined by the set of sample nodes $z_i$ and weights $w_i$. In practice, we use the Clenshaw-Curtis quadrature.  
The corresponding sampling points $(z_{1i},z_{2j})$ are weighted by the product  $w_i w_j$ using the 1D quadrature sample points and weights on both redshift regions with $i=0,\dots, N_{\mrm{z}_1}-1$ and $j=0,\dots,N_{\mrm{z}_2}-1$. Then, one gets the following approximation:
\begin{equation}
C^{\mrm{thick}}_{\ell}(z_1, z_2) \approx  \sum_{i=0}^{N_{\mrm{z}_1}-1}\sum_{j=0}^{N_{\mrm{z}_2}-1} w_i w_j W_1(z_i,z_1)W_2(z_j,z_2) \widehat{P}_\ell(r_i,r_j)
\label{eq-cross-zquadra}
\end{equation}
with the notations $z_i = z_{1i}$, $z_j = z_{2j}$ and  $r_i = r(z_{1i})$, $r_j = r(z_{2j})$ and

\begin{equation}
\widehat{P}_\ell(z_i,z_j) = \int_0^\infty \dx k\ f_\ell(z_i,k) f_\ell(z_j,k)
,\end{equation}
defined with the $f_\ell(z,k)$ function of equation \ref{eq-fell-func}.
 
We use a piecewise integration over a user-defined range $[k_{\mrm{min}}, k_{\mrm{max}}]$ such that 
\begin{equation}
\widehat{P}_\ell(z_i,z_j)  \approx \sum_{p=0}^{N_\mrm{k}-1} I_\ell(k^\ell_p,k^\ell_{p+1};z_i,z_j) 
,\end{equation}
where the $k^\ell_p$ bounds  are related to the roots of $j_{\ell}(x)$ noted $q_{\ell p}$ and the user-defined number of roots $q_{\ell p}$ per sub-interval $[k^\ell_p, k^\ell_{p+1}]$ (see appendix \ref{asec-3Calgo}). 
Then, equation \ref{eq-cross-zquadra} may be rewritten as
\begin{equation}
\begin{split}
C^{\mrm{thick}}_{\ell}(z_1, z_2) \approx \sum_{i=0}^{N_{\mrm{z}_1}}\sum_{j=0}^{N_{\mrm{z}_2}} & w_i w_j W_1(z_i,z_1)W_2(z_j,z_2) \\ &\qquad \times \ \sum_{p=0}^{N_\mrm{k}-1} I_\ell(k^\ell_p,k^\ell_{p+1};z_i,z_j).
\end{split}
\end{equation}
The integral $I_\ell(k^\ell_p,k^\ell_{p+1};r_i,r_j)$ defined as
 
\begin{equation}
I_\ell(k^\ell_p,k^\ell_{p+1};r_i,r_j) = \int_{k^\ell_p}^{k^\ell_{p+1}} \dx k\ f_\ell(k, z_i) f_\ell(k, z_j)
\label{eq-I-integ-cross}
,\end{equation}
 
is computed using the 3C-algorithm of appendix \ref{asec-3Calgo}. Investigating the different steps of the algorithm, one notices that the sampling of the function  $f_\ell(k,z_i)$  along the $k$ axis for a given $[k^\ell_p, k^\ell_{p+1}]$ interval depends on $z_i$ but is independent from $z_j$ and {\it vice versa} for the    $f_\ell(k, z_j)$  function (those samplings are independent from $z_i$). So, one may proceed to $k$-sampling before performing the double sum over $(i,j),$ which is particularly efficient as the  CPU bottleneck is the computation of the spherical Bessel function $j_\ell$. \verb|Angpow| uses a spherical Bessel function implementation based on Numerical Recipes \citep{Press:1992:NRC:148286}. The brute force Cartesian quadrature evolves as $O(N_{\mrm{z}_1} \times N_{\mrm{z}_2}),$ while the optimized version reduces the CPU times scaling  to $O(N_{\mrm{z}_1} + N_{\mrm{z}_2})$. Therefore, as a matter of efficiency, it is more appropriate to exchange the order of the $p$ and $(i,j)$ summations leading to
\begin{equation}
C^{\mrm{obs}}_{\ell}(z_1, z_2) \approx \sum_{p=0}^{N_\mrm{k}-1} \sum_{i=0}^{N_{\mrm{z}_1}}\sum_{j=0}^{N_{\mrm{z}_2}} w_i w_j  W(z_i,z_1)W(z_j,z_2) I_\ell(k_p,k_{p+1};z_i,z_j).
\end{equation}
As a first hint for the 3C-algorithm, we use typically $2^8-2^9$ polynomial approximations for $\ell_\mrm{max}\approx 500-1000$ of the  $f_\ell(k, r_i)$ and $f_\ell(k, r_j)$  functions, $100$ spherical Bessel roots per sub-interval, $99$-point Clenshaw-Curtis quadrature nodes, and weights for $z$ integration in case of $\sigma=0.02$.
\section{Numerical tests}
\label{sec-num-test}
\subsection{Comparison with other codes}
\begin{figure}[t]\centering
\includegraphics[width=1.0\linewidth]{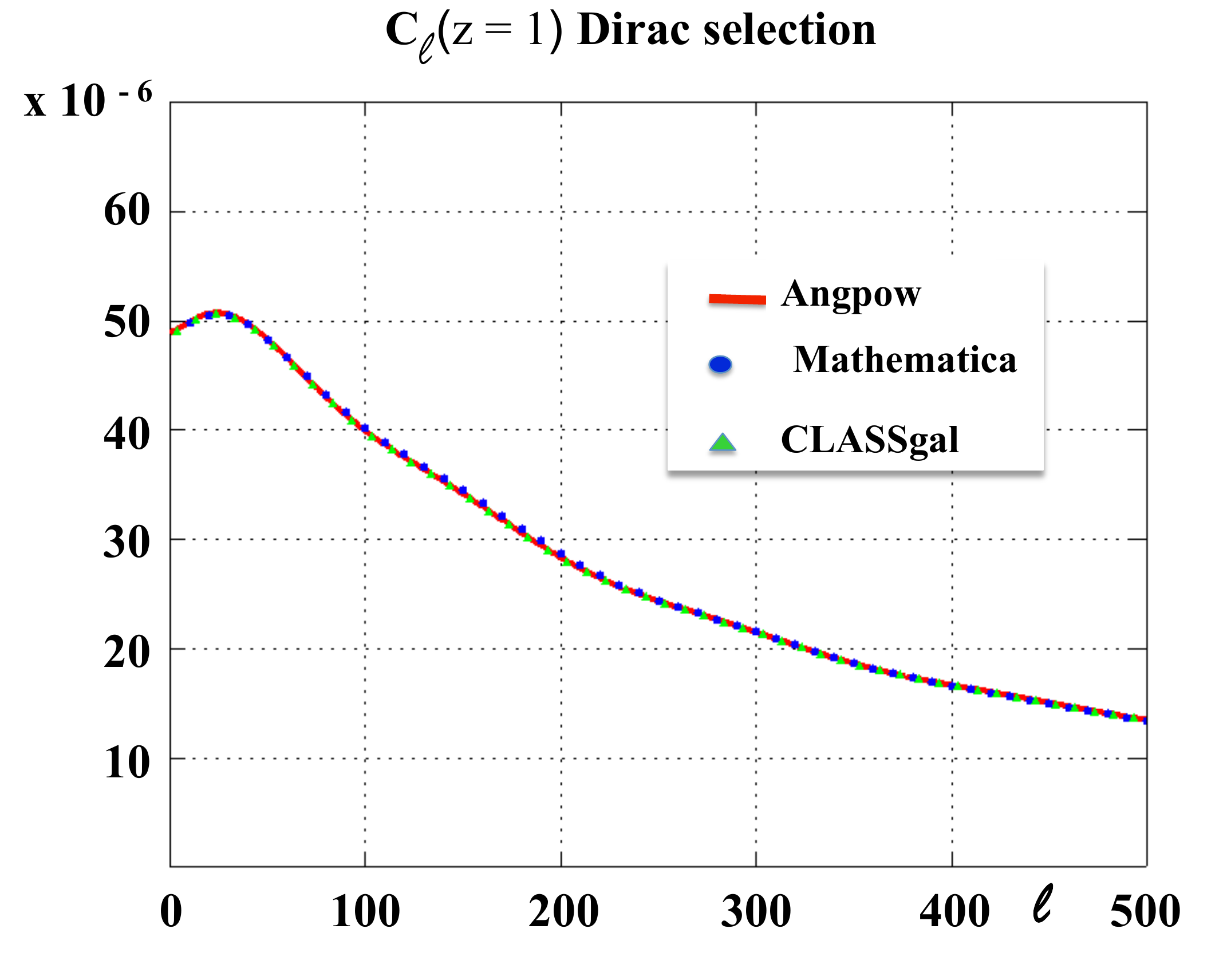}
\caption{Comparison of the computations of the $C_\ell^{\delta}(z=1)$ given by \texttt{Angpow}, \texttt{Mathematica,} and \texttt{CLASSgal} for a Dirac selection function with $k_\mrm{max} = 10\ \mrm{Mpc}^{-1}$.
}
\label{fig-auto-Dirac-M10-class-angpow}
\end{figure}
We proceed now to a numerical comparison of the estimation of $C_\ell(z_1)$ and $C_\ell(z_1,z_2)$ computed by \verb|CLASSgal| \citep{ClassGal} and \verb|Mathematica| \citep{Mathematica11} with Dirac redshift selection functions. Concerning \verb|CLASSgal| (v1.1.3), we have started with the provided  \verb|explanatory.ini|  file where we have modified the cosmological parameters to:  $h=0.679$, $\Omega_{\mrm{b}} =  0.0483$, $\Omega_{\mrm{cdm}} =  0.2582$ and $\Omega_{\mrm{k}} = \Omega_{\mrm{fld}} = 0$. We have also set \verb|k_scalar_max_tau0_over_l_max| to fix the upper bound of the $k$-integration taking into account the maximal $\ell$ value and the redshift mean value. Concerning \verb|Angpow| we have taken advantage of the possibility to read an external file produced by \verb|CLASSgal| as an input $P(k)$ computed at $z=0$ in association to the growth function computed internally given by \citep{1991MNRAS.251..128L, 1992ARA&A..30..499C}.  
Finally, to avoid the Limber's approximation for \verb|CLASSgal|,  we have set the "Limber" threshold much higher than the $\ell$ upper limit.
\begin{figure}[h]\centering
\includegraphics[width=1.0\linewidth]{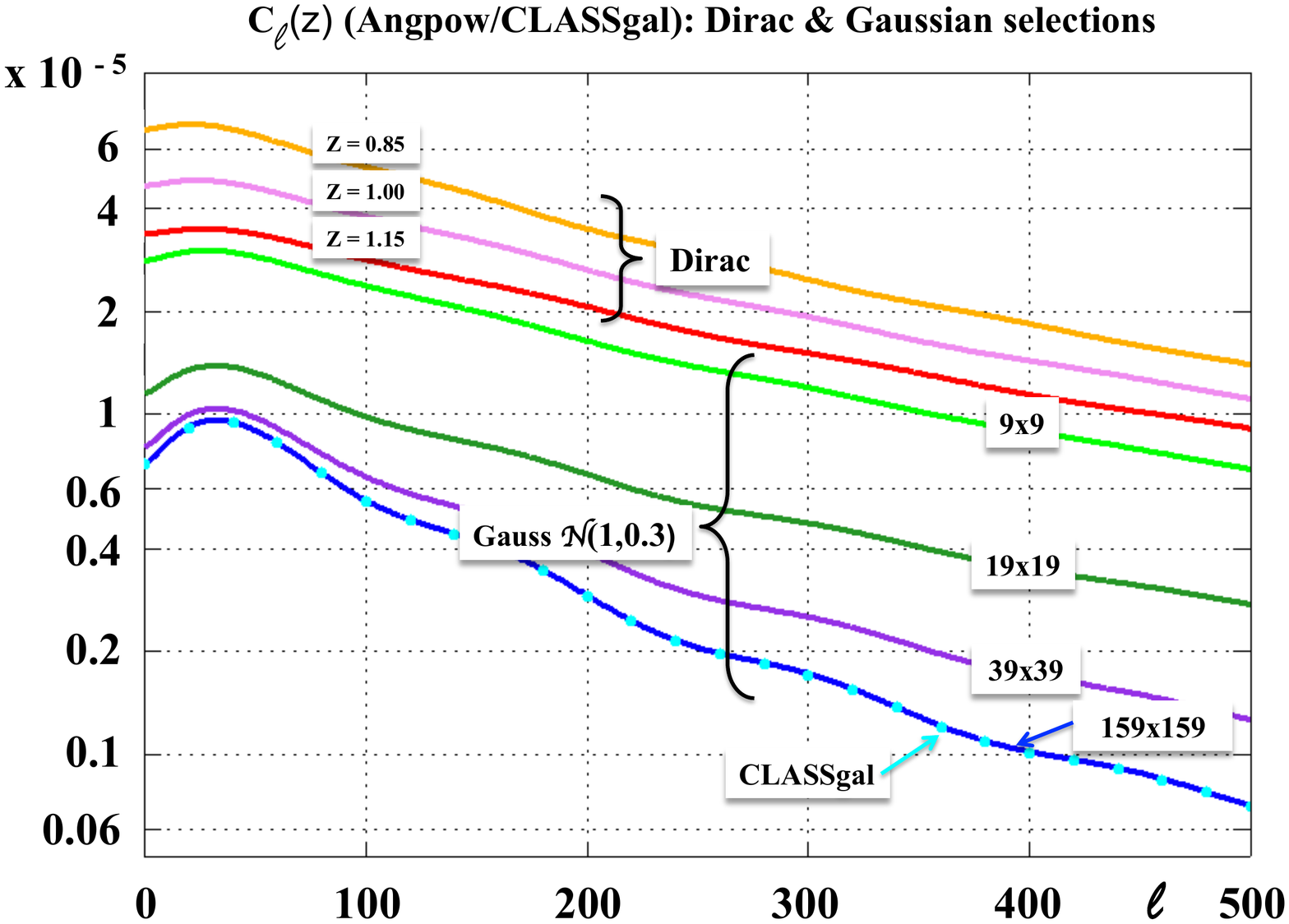}
\caption{Computations of $C_l^{\delta}(z)$ with  Dirac selections centered at $z\in \left\lbrace 0.85, 1.00, 1.15\right\rbrace$ with \texttt{Angpow} alone and  $C_l^{\mrm{thick}}(z)$ with \texttt{Angpow} and \texttt{CLASSgal} using a Gaussian selection function with mean $z=1,$  a width of $\sigma=0.3,$ and a redshift cut at $\pm 5\sigma$ (in all cases $k_\mrm{max}=0.44\ \mrm{Mpc}^{-1}$). For \texttt{Angpow} we use the power spectrum produced by \texttt{CLASSgal} and we have varied the redshift grid sampling resolution from $9\times 9$ to $159\times 159$ points to reach the converged result (blue curve) in good agreement with the \texttt{CLASSgal} result (cyan points). Comparing the $C_l^{\delta}(z=1)$ to the $C_l^{\mrm{thick}}(z=1,\sigma=0.3)$ results we measure the effect of self-cross-correlation inside a thick redshift shell which washes out the matter fluctuation contrast.}
\label{fig-auto-Gauss-class-angpow}
\end{figure}

Results of the auto-correlation $C_\ell^{\delta}(z)$ computations at $z = 1$ using Dirac selection functions are shown in Fig. 1. As the three softwares use the same primordial power spectrum, all the results agree with one other within a maximal relative error of $0.06\%$ on the whole $\ell$ range. 


\begin{figure}
\includegraphics[width=1.0\linewidth]{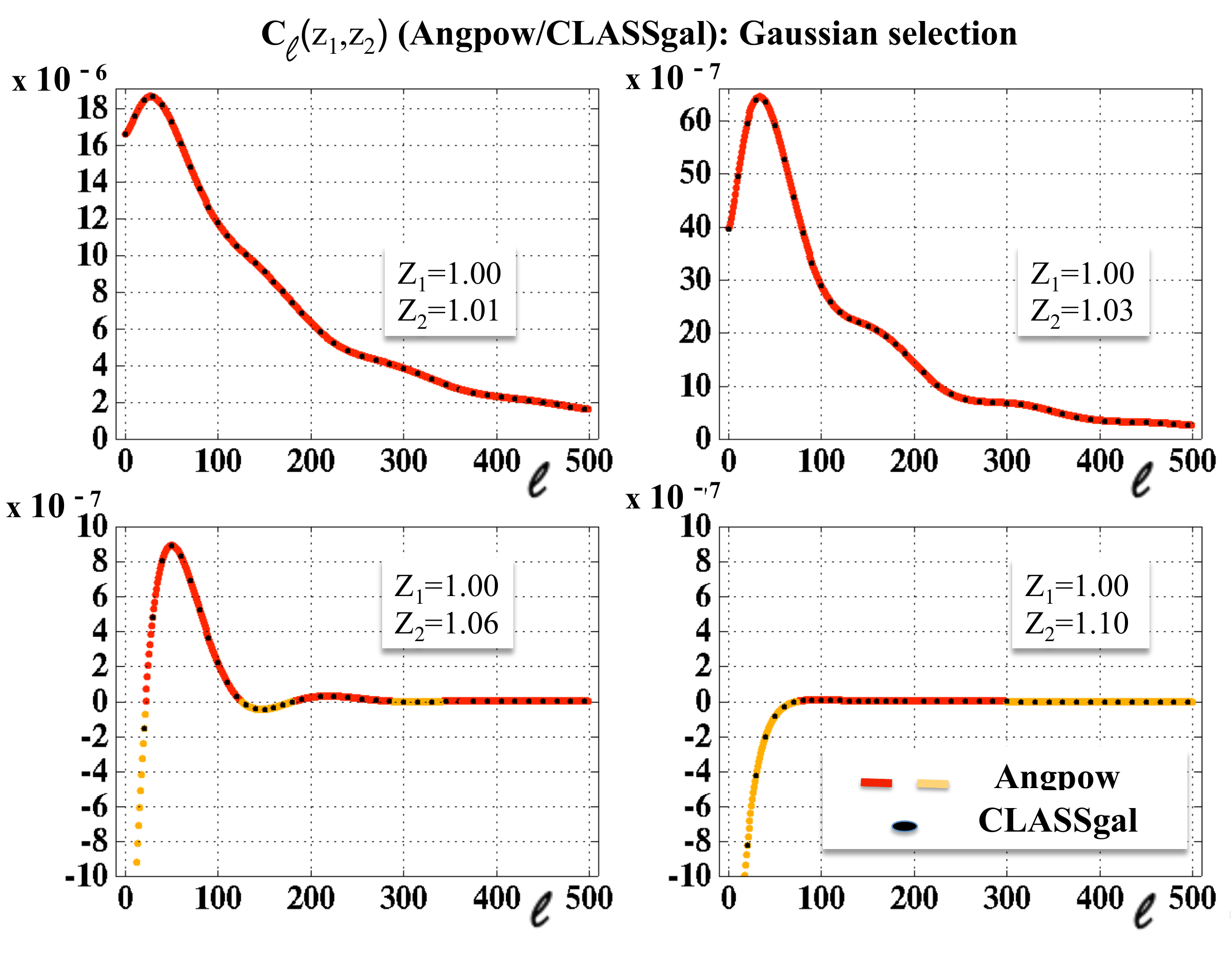}\caption{Comparison between  \texttt{Angpow} (red/orange points) and \texttt{CLASSgal} (black points) for
several cross-correlation spectra with Gaussian  ($\sigma = 0.01$) selection  and $k_\mrm{max}=1\ \mrm{Mpc}^{-1}$. The orange points are used to emphasize negative $C_\ell$.}
\label{fig:Gauss-cross}
\end{figure}

The $C_\ell^{\mrm{thick}}(z)$ auto-correlation  computation within a thick redshift band, selected  by a Gaussian of mean $z=1$ and $\sigma = 0.03$ cut at $\pm 5 \sigma$, has been used as a test bench. But, for this test we have neglected the \verb|Mathematica| software, which is too slow, and have restricted testing to comparison of \verb|Angpow| to \verb|CLASSgal|. Figure \ref{fig-auto-Gauss-class-angpow} shows computation results. The orange, violet, and red curves are produced by \verb|Angpow| using Dirac selection function in the range $\pm 5\sigma$ around the mean redshift $z=1$, while the green, forest green, purple, and blue curves are results of the above mentioned Gaussian selection function but sampled using different grid sizes: $9\times9$, $19\times19$, $39\times39$ and $159\times159$ Clenshaw-Curtis sample points. As the number of points, or equivalently the quadrature order, increases, the $C_\ell^{\mrm{thick}}(z)$ computation converges towards the \verb|CLASSgal| result (cyan points). 
We also address the cross-correlation computations performed by
\verb|Angpow| and \verb|CLASSgal| ($C_\ell^{\mrm{thick}}(z_1,z_2)$)
using Gaussian selection functions ($\sigma=0.01$ and a $\pm
5\sigma$  cut). The results are shown in Figure~\ref{fig:Gauss-cross}.
  One should not be surprised by
    negative values since we are cross-correlating two different
    quantities.  
In both tests  we have used the power spectrum computed at $z=0$ by \verb|CLASSgal| as input to \verb|Angpow|. The agreement  between the two software codes is very  good, keeping the relative residuals at values less than $0.02\%$.
\subsection{Note on Limber's approximation}
\label{seq-limber}
The \verb|Angpow| library can also be used to compute, if desired, the first order Limber's approximation \citep{2008PhRvD..78l3506L}. In such an approximation, the spherical Bessel function is formally reduced to
\begin{equation}
j_\ell(x) \approx \sqrt{\frac{\pi}{2\ell +1}} \delta^D\left(x-\left(\ell+\frac{1}{2}\right)\right).
\label{eq-jlx-Limber}
\end{equation}
In such conditions, the product $k r(z)$ is constrained, and if one uses the following notation for the comoving distance computation with $d_\mrm{H}=c/H_0$ , the Hubble distance ($H_0= 100 h\, \mrm{km}.\mrm{s}^{-1}\mrm{Mpc}^{-1}$ and $c$ the speed of light)  and $E(z),$ the dimensionless Hubble parameter,
\begin{equation}
r(z_\ell(k)) = \frac{l+1/2}{k} = d_\mrm{H}\int_0^{z_\ell(k)} \frac{\dx z}{E(z)}.
\end{equation}
Then, Eq.~\ref{eq-clz1z2-obs} is transformed to the following expression
\begin{equation}
\begin{split}
C_{\ell}^{\mrm{thick}}(z_1, z_2; \sigma_1, \sigma_2) &\approx \frac{2}{d_\mrm{H}^2(2\ell+1)}  
\int_0^\infty \dx k\ W_1(z_\ell(k); z_1, \sigma_1) \\& \times W_2(z_\ell(k); z_2, \sigma_2) E^2(z_\ell(k)) P(k,z_\ell(k)).
\label{eq-Cl-limber}
\end{split}
\end{equation}
This integral can be computed using a divide-and-conquer recursive method with the Gauss-Kronrod 
quadrature \citep{Laurie97calculationof}. The Gauss sample points are a subset of the Gauss-Kronrod sample points and can easily be used to set up an error estimate to drive the recursive algorithm.

\begin{figure}[t]\centering
\includegraphics[width=1.0\linewidth]{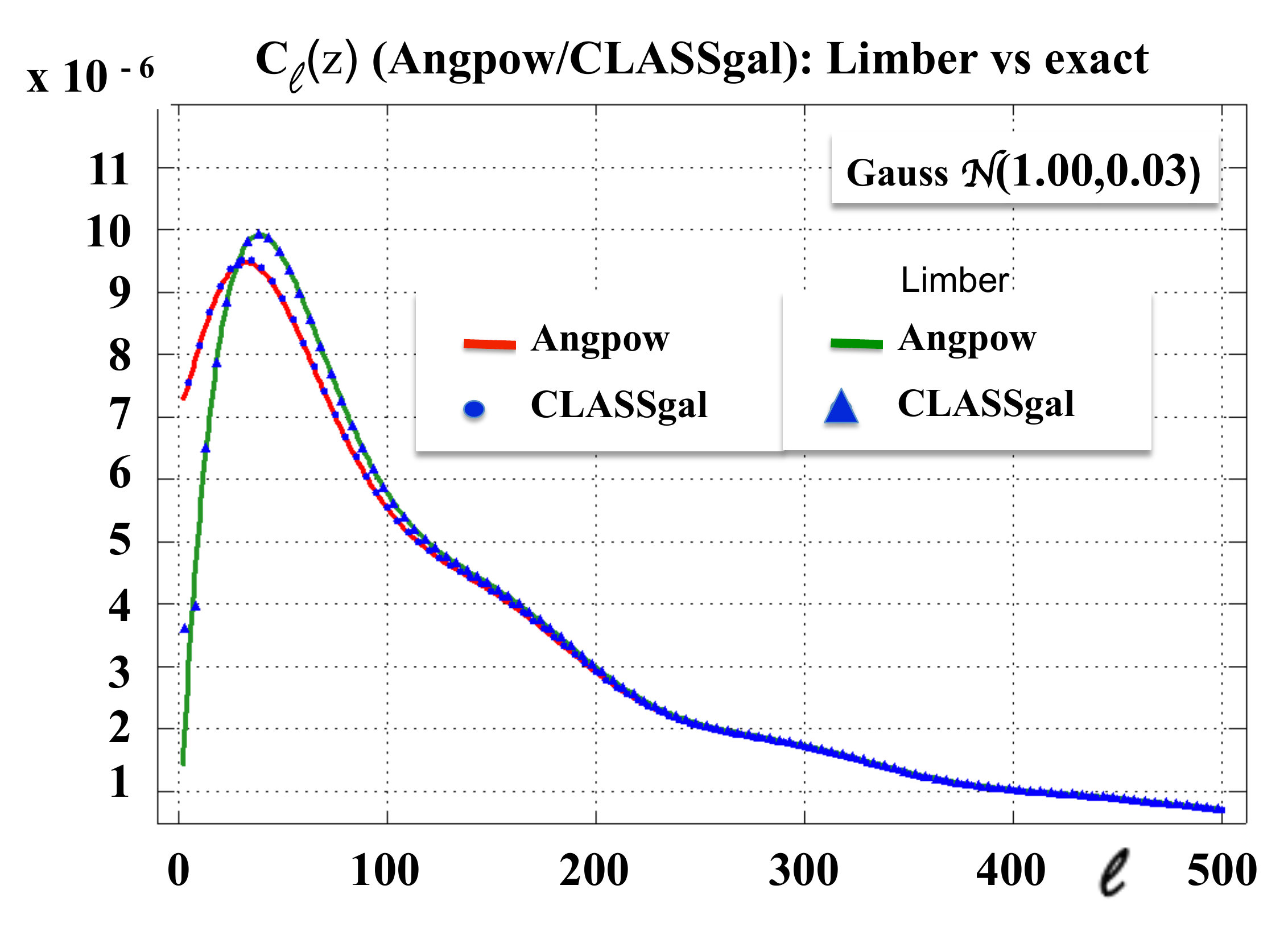}
\caption{Comparison of the computations of the $C_\ell^\mrm{thick}(z)$ given by \texttt{Angpow} and \texttt{CLASSgal} either with the Limber's approximation or the exact computation.}
\label{fig-Cl-angpow-classgal-limber}
\end{figure}

Looking at Eq.~\ref{eq-Cl-limber} one realizes that all the terms are positive, indicating that such approximation
is not suitable for cross-correlation where $C_\ell(z_1, z_2)$ is not guaranteed to be positive as can be explicitly seen in Figure~\ref{fig:Gauss-cross}. We have proceeded to the computation of $C_\ell(z)$ in the case of a Gaussian selection function of mean $z=1$ and $\sigma = 0.03$ for both \verb|Angpow| and \verb|CLASSgal,| with/without the Limber's approximation. The results are shown in Figure~\ref{fig-Cl-angpow-classgal-limber}. The two software codes  agree very well and show that the Limber's approximation can give sizeable errors compared to the exact computation; of the order of the cosmic variance in the given example. So, this Limber's approximation, even if it runs 100 times faster than the exact computation, should then be used with extreme care not only for cross-correlation but also for auto-correlation. 

\begin{figure}[t]\centering
\includegraphics[width=1.0\linewidth]{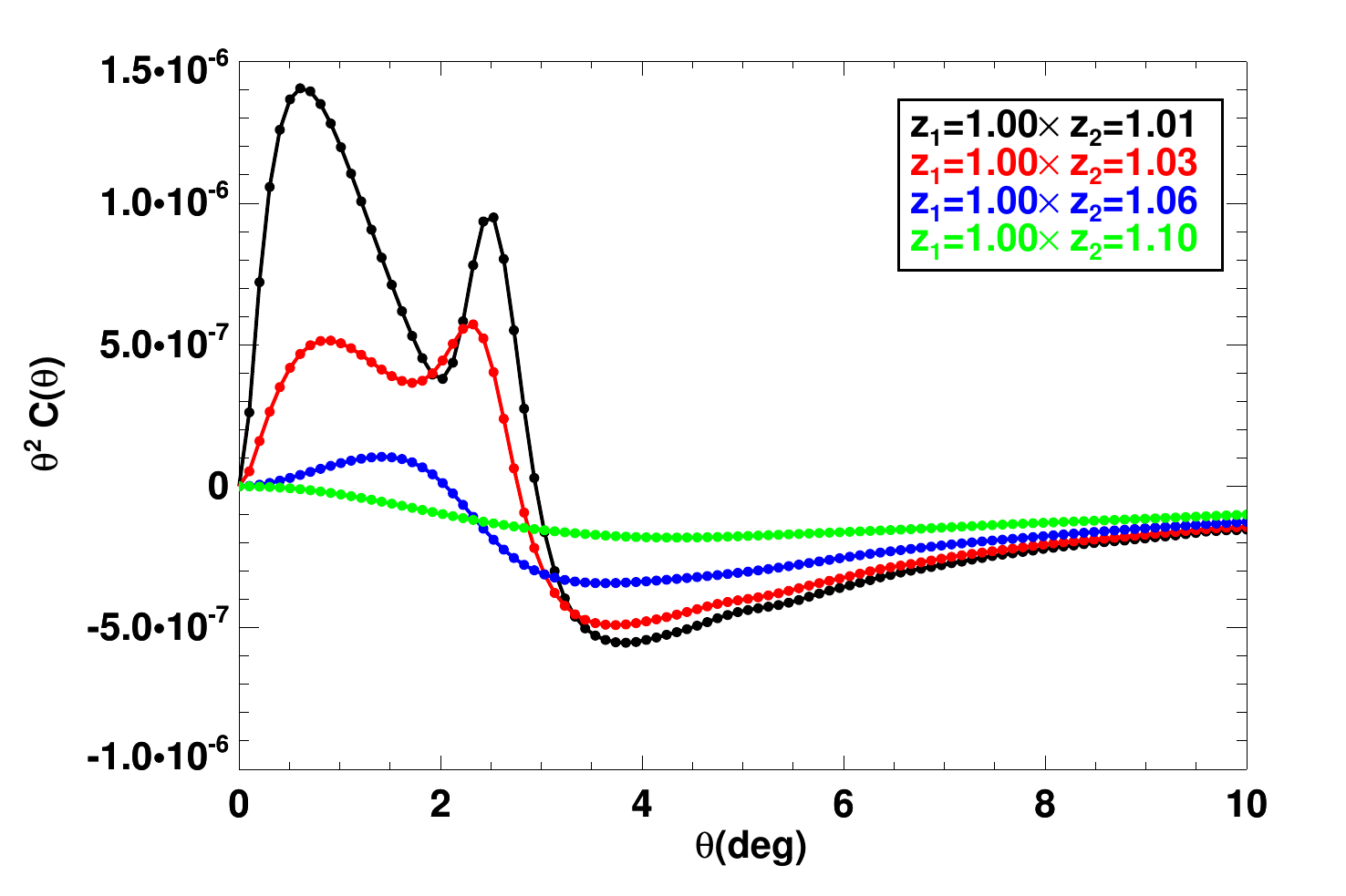}
\caption{Cross-correlations in real space
  corresponding to the spectra shown on
  Figure \ref{fig:Gauss-cross}. The points show where the function
  was evaluated. }
\label{fig:ctheta}
\end{figure}

\subsection{Correlations in real space}

\verb|Angpow| can also quickly compute the correlation
function in real space from the power spectrum 
\begin{equation}
\label{eq:ctheta}
C(\theta; z_1, z_2)=\frac{1}{4\pi} \sum_{\ell=0}^{\ell_\mrm{max}} (2\ell+1)
\tilde{C}_{\ell}(z_1, z_2) P_\ell(\cos \theta)
,\end{equation}
where $P_\ell$ denotes the $\ell-\mathrm{th}$ order Legendre
polynomial. Because the  $C_{\ell}(z_1, z_2)$ values are generally cut at  a given
$\ell_\mrm{max}$, one needs to introduce an apodization to avoid ringing
due to a sharp cut-off. We introduce a Gaussian one (which is the
smoothest in both real and harmonic spaces) as in \citep{ClassGal} so that the $ \tilde{C}_{\ell}(z_1, z_2) $ term in
Eq. \ref{eq:ctheta} is
\begin{equation}
 \tilde{C}_{\ell}(z_1, z_2) = C_{\ell}(z_1, z_2 )e^{-\ell(\ell+1)/\ell_a^2}.
\end{equation}
The apodization length $\ell_a$ may depend on the signal but
for the standard cosmology shown here (around $z=1$) we noticed that using $l_a\simeq
0.4 \ell_{max}$ gives good results. 
Correlations in real space are generally easier to comprehend as is shownin Figure~\ref{fig:ctheta}, which represents the analogue of
Figure~\ref{fig:Gauss-cross} but in real space. Here one may
identify a peak, named the "Baryonic Acoustic Oscillation"
\citep[e.g.,][]{Weinberg2013} that decreases in the cross-correlation
when the distance between shells increases and is finally washed out.

\subsection{Speed tests}
\verb|Angpow| is designed and written in {\tt C++} and parallelization is achieved through OpenMP. To qualify the code we provide four input parameter files and their corresponding outputs obtained in one of our runs. In all tests, we used a $\Lambda$CDM reference cosmology, a $P(k)$ at $z=0$ provided by \verb|CLASSgal|, $\ell_\mrm{max} = 1000$, a 3C-algorithm with $2^9$ Chebyshev polynomial order, and 100 roots per sub-$k$-interval:
\begin{description}
 \item[Test 1]: Auto-correlation with a Dirac selection function at $z=1$ and $k_\mrm{max}=10\ \mrm{Mpc}^{-1}$;
 \item[Test 2]: Cross-correlation with two Dirac selection functions at $z=1$ and $z=1.05$ and $k_\mrm{max}=10\ \mrm{Mpc}^{-1}$;
 \item[Test 3]: Auto-correlation with a Gaussian selection function with ($z_\mrm{mean}=1$, $\sigma_z=0.02$, $5\sigma_z$-cut) and $k_\mrm{max}=1\ \mrm{Mpc}^{-1}$ and a radial quadrature based on $N_\mrm{pts}=139$ sample points;
 \item[Test 4]: Cross-correlation with two Gaussian selection functions with ($z_\mrm{mean,1}=0.90$, $z_\mrm{mean,2}=1.00$) both with  ($\sigma_z=0.02$, $5\sigma_z$-cut) and $k_\mrm{max}=1\ \mrm{Mpc}^{-1}$ and a radial quadrature based on $N_\mrm{pts}=139$ sample points;
\end{description}
We have tested the code both on laptop (Linux, MacOSX) as well as on
Computer Center (CCIN2P3 in France and NERSC in the USA). 
We use OpenMP to distribute the computation of a given $C_\ell$ on a single
thread. Table \ref{table:CCIN2P3-multithread} gives Central Processing Unit (CPU) execution
 times averaged over ten processes. 
  The code wall time  decreases reasonably well with the number of threads
  and a wall time at the $\mathcal{O}$(1s) level can be reached to
reconstruct these accurate spectra.
 Such performances are much higher than those obtained
with {\tt CLASSgal} when not using the Limber approximation.
For example, on our Test-3 setup, running the latter takes
about 15s (on 16 threads), which is to be compared to about 0.5s in Table \ref{table:CCIN2P3-multithread}.
 
\begin{table}
\caption{Wall time (in seconds) measured at CCIN2P3 (on Intel Xeon CPU
  E5-2640 v3 processors) for the test benches
  described in the text, according to the number of OpenMP threads used. Results are
  given for the intel {\tt icpc 15.0} and {\tt gcc 5.2} compilers.}             
\label{table:CCIN2P3-multithread}      
\centering          
\begin{tabular}{c c c c c c}
\hline\hline
\# Threads& 1 & 2 & 4 & 8 & 16 \\ 
\hline
\multicolumn{6}{c}{Linux/icpc}\\
\hline  
Test 1  &0.38   &0.21   &0.13   &0.09   &0.08   \\
Test 2  &0.76   &0.41   &0.23   &0.15   &0.11   \\
Test 3  &3.72   &1.96   &1.05   &0.64   &0.44   \\
Test 4  &9.97   &5.25   &2.79   &1.60   &1.01   \\
\hline
\multicolumn{6}{c}{Linux/gcc}\\
\hline
Test 1  &0.56   &0.30   &0.17   &0.12   &0.09   \\
Test 2  &1.14   &0.60   &0.33   &0.20   &0.14   \\
Test 3  &5.01   &2.59   &1.38   &0.81   &0.50   \\
Test 4  &13.80  &7.07   &3.71   &2.12   &1.27   \\
\hline\hline
\end{tabular}
\end{table}

\begin{figure}[t]\centering
\includegraphics[width=1.0\linewidth]{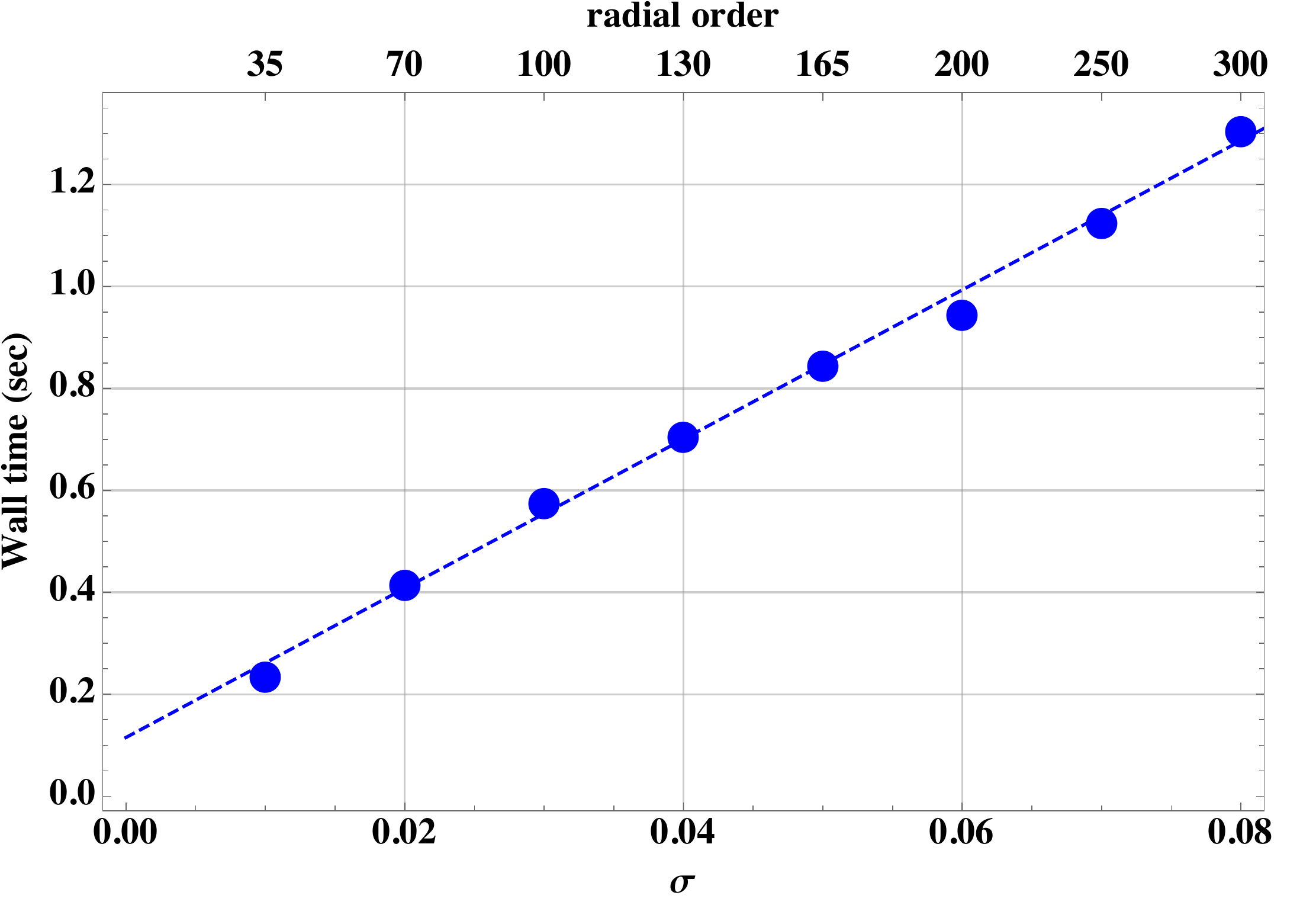}
\caption{Evolution of the wall time with respect to the width of the selection function ($\sigma$) illustrated in the condition of Test 3 and using 16 threads. In the upper scale of the frame is shown an indication of the  \texttt{radial\_order} used to sample the along the $z$ direction for a given $\sigma$ (see section~\ref{sec:code-design}).
}
\label{fig:sig-radord-wt}
\end{figure}
Figure~\ref{fig:sig-radord-wt} shows the dependence of the wall time with respect to the width of the selection function 
($\sigma$) in the  conditions of Test 3 using 16 threads. For a given $\sigma$ , we have chosen the minimal \verb|radial_order| value such that the relative accuracy on the $C_\ell$ is of the order of 0.01\% compared to a computation with a much larger \verb|radial_order| value (see section \ref{sec:code-design}). If one uses a looser criteria on the accuracy of the $C_\ell$ or if the number of sigma is lower than 5, then one may use a lower \verb|radial_order| and gain on the wall time.

\section{Code design and input parameters}
\label{sec:code-design}
\verb|Angpow| is written in C++ which allows both good CPU performances and keeps the code flexible. A front end interface to Python is also foreseen and the code is distributed publicly  at \href{https://gitlab.in2p3.fr/campagne/AngPow}{https://gitlab.in2p3.fr/campagne/AngPow}.  The  \verb|angpow.cc| file is an example of the library usage to perform $C_\ell$ and $C(\theta)$ computations. We also provide  the  \verb|limber.cc| file if one wants to test the Limber's approximation (Sect.~\ref{seq-limber}). The different files are located in self-explained directories: \verb|src|, \verb|inc/Angpow|, \verb|lib|, \verb|data|. Finally, a README file provides details for the installation and build procedures.

The two main classes \verb|Pk2Cl| and \verb|KIntegrator| (located in \verb|angpow_pk2cl.h| and \verb|angpow_kinteg.h| files) are generic codes using abstract base classes. They define interface to the power spectrum function $P(\ell, k, z)$ (class \verb|PowerSpecBase|), the generalisation of $P(k,z)$ used in Eq. ~\ref{eq-clz1z2-obs}; to the comoving distance computation $r(z)$ (class \verb|CosmoCoorBase|); and to the radial/redshift selection functions $W(z)$ (class \verb|RadSelectBase|). The user can derive their own concrete classes to access a third party library or use the ones implemented by default. For instance, we have implemented a file access to a $(k, P(k))$-tuple saved by the \verb|CLASSgal| output. In this implementation, we have coded the growth factor defined in \citep{1991MNRAS.251..128L, 1992ARA&A..30..499C} as minimal  $\tilde{\Delta}_\ell(z,k)$ function  (Eq.~\ref{eq-DeltaFunc-def}) .

To run the executable, one provides an ascii file defining the input
parameters that drive the computational conditions of the algorithm
and define the I/O locations. Some ready-made input parameter files
are also provided (\verb|angpow_bench<n>.ini|) as well as the $C_\ell$
output files (\verb|angpow_bench<n>_cl.txt.REF|) corresponding to the
{\tt icpc} outputs of Table \ref{table:CCIN2P3-multithread}.

Among the different input parameters some are more sensitive than others, as those that deal with the radial/redshift 1D quadrature and the 3C algorithm. Here is a closer look at these parameters:
\begin{itemize}
\item \verb|cl_kmax|: This is the maximal value of $k$ in the $k$-integral. We have not set up an internal algorithm to determine this upper bound. As a hint, one may consider a relation with the factor $\ell_\mrm{max}/r(z_\mrm{min})$. The lower bound on $k$ is internally fixed using the cut-off $x_\mrm{min}(\ell)$ defined as $x<x_\mrm{min}(\ell) \Leftrightarrow j_\ell(x)< \mrm{cut}$ (see the input parameters \verb|jl_xmin_cut| and \verb|Lmax_for_xmin| set as default to $5\ 10^{-10}$ and 2000, respectively). 
\item \verb|radial_order|: If noted $n$, this fixes the number of sample points along one $z$ direction, that is, $N_\mrm{pts} = 2 n -1$ .  The accuracy on the selection function as well as the CPU  increase with $n$ but we keep a $O(n)$ complexity of the $k$-sampling ( see Figure~\ref{fig:sig-radord-wt} );
 \item \verb|chebyshev_order|: If noted $N$, this fixes the degree $d$ of the Chebyshev polynomial approximation of the $f_\ell(k, z_i)$ and $f_\ell(k, z_j)$ functions (Eq.~\ref{eq-fell-func}); that is, $d=2^N$. Keeping the same degree of approximation for both functions guarantees a power of 2 for the product approximation. Even if  not mandatory, this helps in getting CPU performance for the DCT-I transform (using the \verb|FFTW| library). When $\ell_\mrm{max}$ increases it may be worth updating this parameter by 1 unit step. For $\ell_\mrm{max}=500$ \verb|chebyshev_order| is set to 8 by default. Increasing the angular spectrum computation up to $\ell_\mrm{max}=1000$ keeping this default value leads to absolute error of the order of $10^{-6}$ for Tests 1 and 2 and $5\ 10^{-10}$ for Test 3 and 4, then to get better accuracy in this case we use  $\mathtt{chebyshev\_order} = 9$.
 \item \verb|n_bessel_roots_per_interval|: This is the number of Bessel roots $q_{\ell p}$ used to define the bounds of the integral $I_\ell(k^\ell_p,k^\ell_{p+1};r_i,r_j)$ (Eq.~\ref{eq-I-integ-cross}). By default it is set to 100. There is an interplay with the \verb|chebyshev_order| parameter as a lower \verb|n_bessel_roots_per_interval| value is coherent with a lower \verb|chebyshev_order|.
 \item \verb|total_weight_cut| and \verb|deltaR_cut|: These two threshold parameters are used to avoid unnecessary 3C algorithm processing (especially for the $k$-sampling of the $f_\ell(k, z_i)$ or $f_\ell(k, z_j)$ functions). So, we do not consider a couple $(z_i, z_j)$ for which either the product $w_i w_j  W(z_i,z_1)W(z_j,z_2)$ is too low (\verb|total_weight_cut| cut) or the radial distance $|r(z_i)-r(z_j)|$ is too large to produce a sizeable contribution to the final $C_\ell$. The \verb|deltaR_cut| cut is in Mpc units and is used in conjunction with \verb|has_deltaR_cut| set to 1. These two threshold parameters depend on the use case under consideration and for preliminary tests we recommend to set both \verb|total_weight_cut|  and \verb|has_deltaR_cut| to 0.
\end{itemize}
\section{Summary and outlooks}
\label{sec-summary}
We have set up a fast and generic software to compute the tomographic $C_\ell(z_1,z_2)$ with redshift selection functions.  \verb|Angpow| is versatile enough to accept user-defined matter power spectrum $P(k)$, transfer functions $\tilde{\Delta}_\ell(k,z),$ and cosmology. The code provides an accurate computation of the auto and cross-correlation power spectra, checked by comparison with other codes, which is fast enough to be included inside MCMC cosmology softwares. The rapidity of the software relies on the use of the 3C-algorithm, adapted to the computation of integrals over spherical Bessel functions, while other codes rely on the Limber's approximation. We emphasize that the Limber's approximation can lead to incorrect $C_\ell(z_1,z_2),$ especially in the case of cross-correlations, as Limber's Dirac functions cannot model the interferences between two spherical Bessel functions at different redshifts.

This code is thus fast and accurate enough to be used to test cosmological parameters, and perform tomographic analysis of the galaxy distribution. The definition of the $\tilde{\Delta}_\ell(k,z)$ function is general and can accept zero order function as $\tilde{\Delta}^{\rm mat}_\ell(k,z)$, but also relativistic corrections such as the redshift space distortions or the gravitational lensing; despite these corrections, it can also contain spherical Bessel functions \citep[see e.g.,][]{ClassGal}. Because \verb|Angpow| provides the correct angular power spectra for cross-correlations, it can be a key tool to perform an integrated approach to cosmology, as advertised in \citep{Nicola2016}. In particular, we propose this tool for deriving the auto- and cross-correlation angular power spectra for galaxy clustering, but also with angular power spectra from cosmic shear and the cosmological microwave background.

\verb|Angpow| is publicly available\footnote{from \href{https://gitlab.in2p3.fr/campagne/AngPow}{https://gitlab.in2p3.fr/campagne/AngPow}.} and can be interfaced to other codes; a Python interface is foreseen. At the moment the code only accepts two redshift bins but soon it will be generalized to any number of bins. Feedback from \verb|Angpow| users would be greatly accepted.

\begin{acknowledgements}
The authors want to thank M. Reinecke who kindly provided pieces  of the code, and J. D. McEwen for fruitful discussion on the Chebyshev transform.
\end{acknowledgements}
\bibliographystyle{aa}
\bibliography{Angpow}
\appendix
\section{Clenshaw-Curtis-Chebyshev algorithm (3C-algorithm)}
\label{asec-3Calgo}
Each integral type of  Eq.~\ref{eq-I-integ-cross} involves the product of "highly" oscillatory functions.  The purpose of this section is not to provide a review of all the integration methods used in the different fields of physics to tackle such a difficult task.. To focus on our case, where we have to deal with (at least) the product of spherical Bessel functions, the authors point out that the reader may find either specific integral solving rules as in \citep{1993math......7213G} or general methods as in  \citep{LUCAS1995217}. However we need a precise and  also very fast method. We cannot rely on methods that solve the problem of a product of spherical Bessel functions multiplied by a regular function. In fact, both the primordial power spectrum and the extension beyond the matter density $\tilde{\Delta}^{\mrm{mat.}}(z,k)$ may show oscillation features in the form of derivative of spherical Bessel functions. So, we have searched and found  a general method that meets our requirements of precision and speed.

Eq.~\ref{eq-I-integ-cross}  is a special case of the following generic integral after a proper change of variable
\begin{equation}
I = \int_{-1}^1 \dx x f(x) g(x).
\end{equation}
To get an approximate value of this integral, we use in this section the Clenshaw-Curtis quadrature at order $N_{\mrm{cc}}$ (noting $h=f \times g$):
\begin{equation}
I \approx \sum_{k=0}^{N_{\mrm{cc}}} w_k f(x_k)g_(x_k) = \sum_{k=0}^{N_{\mrm{cc}}} w_k h(x_k)
\label{eq-CC-productInt}
,\end{equation}
where the sampling points are defined as $x_k = \cos k\pi/N_{\mrm{cc}}$ ($k=0,\dots,N_{\mrm{cc}}$) and the corresponding weights $w_k$ are addressed later in the section. But, if the functions $f$ and $g$ have a  highly oscillatory behavior, one needs, in principle, to use large values of $N_{\mrm{cc}}$ to reach a sufficient accuracy level. In that case, dealing with the above sum may not be computationally efficient. The idea is then to use Chebyshev series to approximate both functions $f$ and $g$. Then,  one performs the product of both Chebyshev series, which is also a   Chebyshev series but with a higher order, and finally one uses a fast Clenshaw-Curtis weights computation to perform the last weighted sum. We briefly describe those steps leaving the details of the demonstration that the interested reader can find in \citep{BaszenskiTasche1997a}.

Let $f_N$ be a  polynomial approximation of $f$ of degree $N-1$. We expend $f_N$ onto the following basis of the first kind of Chebyshev polynomials  $\{T_n; n=0,\dots,N-1\}$  which have the property $T_n(\cos\theta)=\cos n\theta$:
\begin{eqnarray}
f_N(x) = \frac{a_0}{2}+\sum_{k=1}^{N-1} a_k T_k(x).
\end{eqnarray}
To determine the $a_k$ coefficients one uses the following sampling vector
\begin{equation}
\mbb{f}^{(N)} =( f( t_\mu^{(N)}))^T\qquad \mrm{with} \qquad t_\mu^{(N)} \equiv \cos \frac{\mu \pi}{N};\; \mu=0,\dots,N
,\end{equation} 
 of length  $N+1$ and related to the vector $\mbb{a}^{(N)} = (a_0,\dots,a_{N-1},0)^T$ by the linear algebra relation
\begin{equation}
\mbb{a}^{(N)} = \frac{2}{N}\mbb{C}^I_N\, \mbb{f}^{(N)}
\label{eq-aCoeff}
,\end{equation}
with $\mbb{C}^I_N$ being a discrete cosine transform of type-I (DCT-I) matrix of dimension $(N+1)^2$. Similarly, we note $g_M$ a polynomial approximation of $g$ of degree $M-1$ from which we determine  the sampling vector  $\mbb{g}^{(M)}$ using the sample points $t_\mu^{(M)}$. The coefficient vector $\mbb{b}^{(M)} = (b_0,\dots,b_{M-1},0)^T$ is related to $\mbb{g}^{(M)}$ using a relation similar to  Eq.~\ref{eq-aCoeff}, namely
\begin{equation}
\mbb{b}^{(M)} = \frac{2}{M}\mbb{C}^I_M\, \mbb{g}^{(M)}.
\label{eq-bCoeff}
\end{equation}
By combining the polynomial approximations $f_N$ and $g_M$, the function $h$ is then approximated by a Chebyshev series of degree $N+M-2$ with coefficient vector $\mbb{c}^{(P)}$ of length $P+1$ with $P\geq N+M-1$.  Using a relation of type Eq.~\ref{eq-aCoeff}, the vector  $\mbb{c}^{(P)}$ is related to the sampling vector
\begin{equation}
\mbb{h}^{(P)} = (h(t_\mu^{(P)}))^T;\; \mu=0,\dots,P.
\end{equation}
To get $\mbb{h}^{(P)}$ it is not necessary to compute $\mbb{c}^{(P)}$ and proceed to an inversion of a DCT-I matrix. The key point is that if we note $\odot,$ the component-wise multiplication, one has
\begin{equation}
\mbb{h}^{(P)}  = \mbb{f}^{(P)} \odot \mbb{g}^{(P)}  
.\end{equation}
Moreover, $\mbb{f}^{(P)}$ ($\mbb{g}^{(P)}$) is obtained from $\mbb{a}^{(N)}$  ($\mbb{b}^{(M)}$) of length $N+1$ ($M+1$) by an extension to a larger vector  at least of length $N+M$ noted $\tilde{\mbb{a}}^{(P)}$ ($\tilde{\mbb{b}}^{(P)}$) by appending with zeros:
\begin{eqnarray}
\tilde{\mbb{a}}^{(P)} &=& (\mbb{a}^{(N)}, 0, \dots, 0), \nonumber \\
\tilde{\mbb{b}}^{(P)} &=& (\mbb{b}^{(M)}, 0, \dots, 0).
\end{eqnarray}
Then, the sampling vector used in Eq.~\ref{eq-CC-productInt} where one identifies $N_{\mrm{cc}}=P$ is determined by
\begin{equation}
\mbb{h}^{(N_{\mrm{cc}})} = (\mbb{C}^I_{N_{\mrm{cc}}}\, \tilde{\mbb{a}}^{(N_{\mrm{cc}})}) \odot  (\mbb{C}^I_{N_{\mrm{cc}}}\, \tilde{\mbb{b}}^{(N_{\mrm{cc}})})
\label{eq-hSample}
,\end{equation}
using $\mbb{C}^I_{N_{\mrm{cc}}}$ the DCT-I matrix of dimension $(N_{\mrm{cc}}+1)^2$ and the inversion property $(\mbb{C}^I_P)^{-1}=(2/P) \mbb{C}^I_P$. In some sense, for both $f$ and $g$ approximation sampling vectors, we have performed a Chebyshev basis change to a larger parameter space compatible with the polynomial degree involved in the product $f^{(N)} \times g^{(M)}$.

The second key point is that the Clenshaw-Curtis weights associated to $\mbb{h}^{(N_{\mrm{cc}})}$ in Eq.~\ref{eq-CC-productInt} can also be  computed with a DCT-I transform from the vector $(2/N_{\mrm{cc}})(1, 0, -1/3, 0, -1/15, \dots, ((-1)^k+1)/2(1-k^2),\dots)$ of length $N_{cc}+1$ \citep{Waldvogel2006} (the normalization depends on the exact definition of the DCT-I used). 

So, to perform the integral given by Eq.~\ref{eq-CC-productInt}, one needs $4+1$ DCT-I transforms,  1 for the Clenshaw-Curtis weights and 4 to transform the Chebyshev coefficients vectors. The DCT-I transform may be implemented using $O(n \log n)$ efficient algorithm, for example, the FFTW library \citep{FFTW05} used by \verb|Angpow|.  \verb|Angpow| uses  a power of $2$ for $N$, $M,$ and also $P$  (keeping $P\geq N+M-1$) for a fast implementation. The 3C-algorithm is a special case of a more general class of algorithms dealing with the product of polynomials. We note that  according to reference \citep{DBLP:journals/corr/abs-1009-4597} an even faster algorithm (although moderate)  might be implemented in a future version of \verb|Angpow| if necessary. We note finally that this general method can be applied to use cases beyond the power spectrum computation in other fields of interest.
\end{document}